\documentclass[iop]{emulateapj}

\usepackage{amsmath}
\usepackage{graphicx,epsfig,epsf,graphics,epstopdf}
\usepackage{natbib,hyperref}
\def\apj{ApJ}
\def\nat{Nature}

\def\apjl{ApJL}
\def\apjs{ApJS}

\def\aap{A\&A}

\def\sun{\odot}

\def\teff{T_\mathrm{eff}}
\def\tsurf{T_\mathrm{surf}}

\def\Eeff{E_\mathrm{eff}}

\shorttitle{UV Surface Environment of Earth-like Planets Orbiting FGKM Stars Through Geological Evolution}
\shortauthors{Rugheimer et al.}

\begin{document}


\title{UV Surface Environment of Earth-like Planets Orbiting FGKM Stars Through Geological Evolution}


\author{S. Rugheimer\altaffilmark{1,3}, A. Segura\altaffilmark{2}, L. Kaltenegger\altaffilmark{3} and D. Sasselov\altaffilmark{1}}

\submitted{This has appeared in the literature as S. Rugheimer et al. 2015 ApJ, 806, 137. doi:10.1088/0004-637X/806/1/137}





\altaffiltext{1}{Harvard Smithsonian Center for Astrophysics, 60 Garden st., 02138 MA Cambridge, USA}
\altaffiltext{2}{Instituto de Ciencias Nucleares, Universidad Nacional Aut\'onoma de M\'exico, M\'exico}
\altaffiltext{3}{Carl Sagan Institute, Cornell University, Ithaca, NY 14853 USA}



\begin{abstract}

The UV environment of a host star affects the photochemistry in the atmosphere, and ultimately the surface UV environment for terrestrial planets and therefore the conditions for the origin and evolution of life. We model the surface UV radiation environment for Earth-sized planets orbiting FGKM stars at the 1AU equivalent distance for Earth through its geological evolution. We explore four different types of atmospheres corresponding to an early Earth atmosphere at 3.9 Gyr ago and three atmospheres covering the rise of oxygen to present day levels at 2.0 Gyr ago, 0.8 Gyr ago and modern Earth \citep[following][]{kaltenegger2007}. In addition to calculating the UV flux on the surface of the planet, we model the biologically effective irradiance, using DNA damage as a proxy for biological damage. We find that a pre-biotic Earth (3.9 Gyr ago) orbiting an F0V star receives 6 times the biologically effective radiation as around the early Sun and 3520 times the modern Earth-Sun levels. A pre-biotic Earth orbiting GJ 581 (M3.5V) receives 300 times less biologically effective radiation, about 2 times modern Earth-Sun levels. The UV fluxes calculated here provide a grid of model UV environments during the evolution of an Earth-like planet orbiting a range of stars. These models can be used as inputs into photo-biological experiments and for pre-biotic chemistry and early life evolution experiments.

\end{abstract}



\section{INTRODUCTION}

\renewcommand{\arraystretch}{0.6}

Thousands of  extrasolar planets have been found to date with thousands more awaiting confirmation from space and ground-based searches. Several of these planets have been found in or near the circumstellar Habitable Zone \citep[see e.g.][]{quintana2014, borucki2013, kaltenegger2013, batalha2013, borucki2011, kaltenegger2011, udry2007} with masses and radii consistent with rocky planet models. The quest for finding habitable planets focuses on main sequence stars with lifetimes conducive for the origin and evolution of life, namely the F, G, K and M spectral types with stellar main sequence lifetimes at least 2 Gyr or more. The UV environment of a host star will affect the atmosphere, and ultimately the surface UV environment for terrestrial planets. \citep[see e.g.][]{sato2014, rugheimer2014, rugheimer2013, segura2003}. 

Future mission concepts to characterize Earth-like planets are designed to take spectra of extrasolar planets with the ultimate goal of remotely detecting atmospheric signatures that can indicate habitability and life \citep[e.g.][]{beichman1999, beichman2006, cash2006, traub2006, kaltenegger2006, seager2015}.The UV surface environment for planets is an important component in providing boundary conditions for biological models exploring the origin of life on Earth as well as planets orbiting other types of stars.

Depending on the intensity, UV radiation can be both useful and harmful to life as we know it. UV radiation from 180 - 300 nm can inhibit photosynthesis and cause damage to DNA and other macromolecule damage \citep{kerwin2007, tevini1993, matsunaga1991, voet1963}. However, these same wavelengths also drive several reactions thought necessary for the origin of life \citep[e.g.][]{senanayake2006, barks2010, sutherland2012, sutherland2015}. In this paper we model the UV surface radiation environments for both pre-biotic and post-biotic planets orbiting other stars at the 1AU equivalent distance based on Earth's evolution.

Previous studies used an analytic atmospheric attenuation model to model UV environment for Archean Earth \citep{cnossen2007, cockell2000a, cockell1999, cockell1998} as well as F stars \citep{sato2014}. An earlier study \citep{segura2003} examined the amount of UV radiation reaching the surface of an Earth-like planet with varying oxygen atmospheric concentration orbiting an F2V, G2V (the Sun), and K2V host star. Our paper expands the grid of host stars as well as models of the UV surface environments for atmospheres that correspond to geological epochs throughout Earth's evolution \citep[following][]{kaltenegger2007}. We focus on four geological epochs corresponding to a 3.9 Ga\footnote{Ga - billion years ago} (assumed prebiotic), 2.0 Ga, 0.8 Ga and the modern Earth atmosphere for a grid of FGKM host stars.


In Section 2, we describe our model, section 3 presents the calculated UV fluxes at the surface and top of the atmosphere of an Earth-like planet for 12 stellar types and the 4 different atmosphere models through geological time. In Section 4 we conclude by summarizing the results and discussing their implications.

\section{MODEL DESCRIPTION}

\subsection{Stellar and Planetary Model}

We use a grid of host stars from F0V to M8V ($\teff$ = 7000K to 2400K) \citep[see][]{rugheimer2013, rugheimer2014}. All stellar models use observations in the UV by the IUE\footnote{http://archive.stsci.edu/iue} except for 3 M dwarfs with HST data and reconstructed Ly-$\alpha$ fluxes \citep{france2013} up to 3000 \AA\  combined with PHOENIX \citep{allard2014, allard2000} or ATLAS \citep{kurucz1979} stellar models\footnote{BT-Settl PHOENIX models are used for the M stars and ATLAS models are used for the FGK stars} for larger wavelengths (F0V, F7V, Sun (G2V), G8V, K2V, K7V, M1V, M3V, M8V\footnote{The M1V, M3V and M8V are the active stellar models defined in \citet{rugheimer2014}.}) The three M dwarfs with updated UV data are, GJ 581 (M3V, $\teff$ = 3498K), GJ 832  (M1.5V, $\teff$ = 3620K), and GJ 1214 (M4.5V, $\teff$ = 3250K) \citep{france2013}. 

To model the planetary atmospheres, we use a coupled 1D radiative-convective atmosphere code developed for rocky exoplanets. It iterates between a 1D climate \citep{kasting1986, pavlov2000, haqq2008} and a 1D photochemistry code \citep{pavlov2002, segura2005, segura2007} to calculate the atmosphere transmission of UV fluxes to the ground of Earth-sized planets.

We simulate the effects of stellar radiation on a planetary environment with an altitude range that extends upwards to 60km, corresponding to a pressure of 1mbar, with 100 height layers. A two-stream approximation \citep[see][]{toon1989}, which includes multiple scattering by atmospheric gases, is used in the visible/near IR to calculate the shortwave fluxes. Four-term, correlated-k coefficients parameterize the absorption by O$_3$, H$_2$O, O$_2$, and CH$_4$ \citep{pavlov2000}. Clouds are not explicitly calculated. Clouds can either reduce or enhance UV radiation reaching the surface of an Earth-like planet  \cite{grant1997, parisi2004}. The climatic effects of clouds on the temperature vs. pressure profile are included by adjusting the planet's surface albedo to the value that for the modern Earth-Sun system yields a surface temperature of 288K \citep[following][]{kasting1984, pavlov2000, segura2003, segura2005}. The photochemistry code, originally developed by \citet{kasting1985} solves for 55 chemical species linked by 220 reactions using a reverse-Euler method \cite[see][and references therein]{segura2010}. 

For the geological epoch at 3.9 Ga, we use a 1D photochemical model for high-CO$_2$/high-CH$_4$ terrestrial atmospheres \citep[see][and references therein]{pavlov2001, kharecha2005, segura2007}. This model simulates an anoxic atmosphere composed of 0.9 bar of N$_2$ and fixed amounts of CO$_2$ and CH$_4$. We run the radiative-convective model to convergence and then use the resulting temperature profile to run the photochemical model that contains 73 chemical species involved in 359 reactions. The model spans the region from the planetary surface up to 64 km in 1-km steps.  All of the simulated planets at 3.9 Ga are assumed to be devoid of life; hence, none of the compounds in the atmosphere are considered to have a biological source. 


\subsection{Simulation Set-Up}

We focus on four geological epochs from Earth's history to model the UV environment on the surface of an Earth-like planet at the 1 AU equivalent distance from its host star. The geological evidence from 2.8 - 3.5 Ga is consistent with an atmosphere with similar atmospheric pressure as modern Earth \citep{som2012, marty2013}. Therefore, we use one bar for the surface pressure for all epochs. 

\begin{figure*}[ht!]
\centering
\includegraphics[scale=0.75,angle=0]{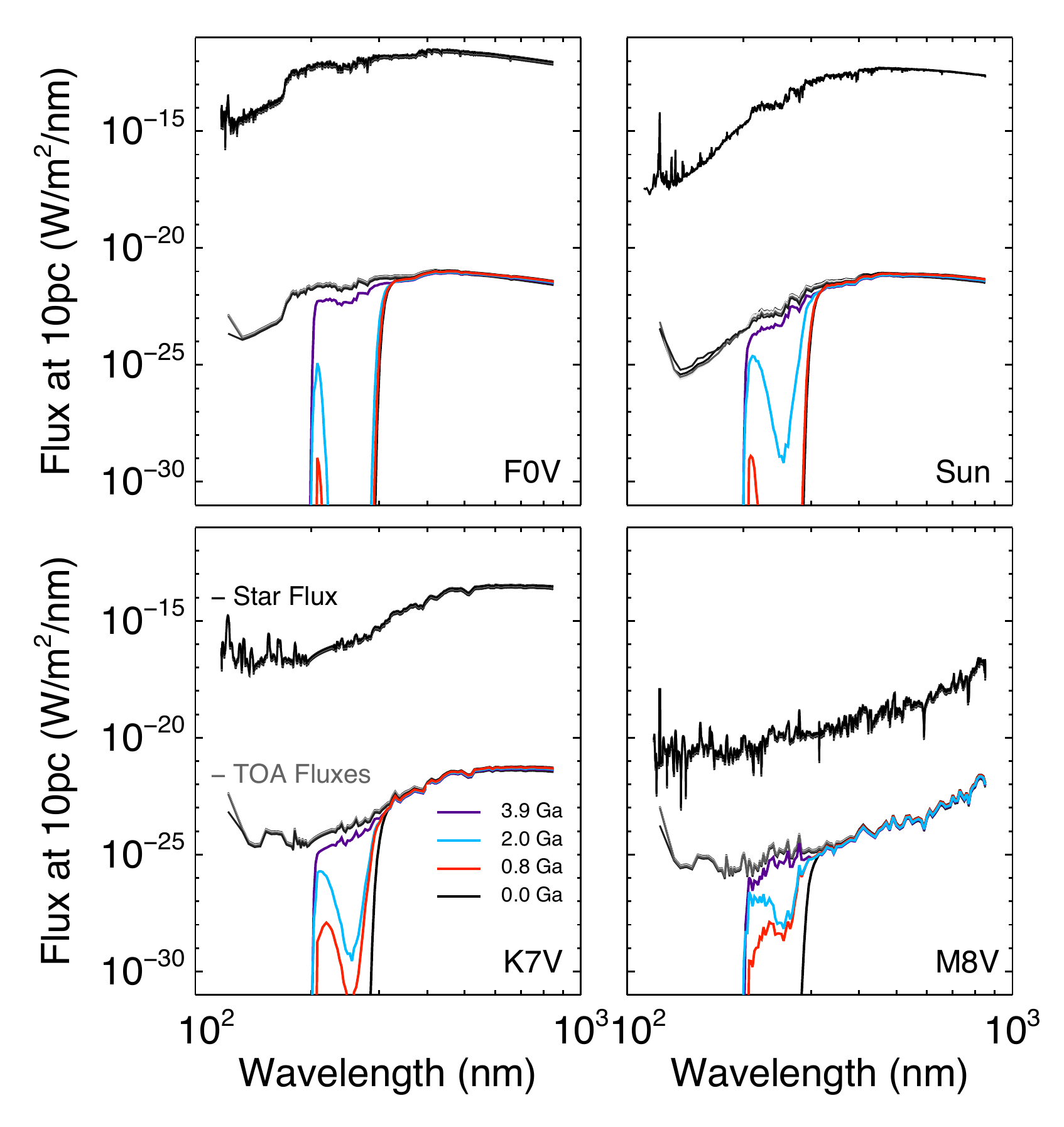}
\caption{The stellar flux (black lines), TOA fluxes (gray lines) and surface UV fluxes (colored lines) for each of the four geological epochs for an F0V (top left), the Sun (top right), a K7V (bottom left) and M8V (bottom right) host star. \label{StarGround}}
\end{figure*}

Currently no model can predict the rate of the evolution of life on planets around different stars which themselves have a different main sequence lifetimes. For the Earth-Sun case, we use a solar evolution model for each epoch \citep{claire2012}. We reduce the stellar flux of all other host stars by the same factor as a first order approximation of how much stellar flux would be received at a corresponding epoch for the other host stars. This procedure is not meant to capture the nuances of stellar evolution. Rather, it is intended to compare across different stellar hosts for planets that receive the same bolometric flux with different atmospheric compositions modeled after Earth's evolution. This translates into different evolutionary stages of the star-planet system as well as into different distances from the the host stars corresponding to 1.15 AU - 1 AU (75-100\% current insolation). Especially for M stars, which evolve much more slowly on the main sequence, reducing the stellar flux by the same amount as Earth is equivalent to increasing its distance from the host star. We model each of the four atmospheres for planets orbiting stars with spectral types of F0V, F7V, the Sun (G2V), G8V, K2V, K7V, M1V, M3V, and M8V to sample the FGKM spectral range along with three observed MUSCLES stars, GJ 581, GJ 832, and GJ 1214 with well characterized UV observations from HST.

The first epoch corresponds to a pre-biotic world, similar to early Earth at 3.9 Ga with a CO$_2$ dominated atmosphere. We model the atmosphere with a fixed surface mixing ratio of CO$_2$ = 0.1 and CH$_4$ =  $1.65 \times 10^{-6}$ and no biological gas fluxes. We assume a stellar flux of 74.6\% of the present day Earth-Sun system. 

The next three epochs represent planets with an active biosphere with oxygenic photosynthesis. To model atmospheres corresponding to a similar stage of biological evolution as Earth's, we use calculated biological surface fluxes from the Earth-Sun model as inputs for the same epoch for all other stellar types as detailed in the following three paragraphs. 

The second epoch corresponds to 2.0 Ga, when oxygen starts to build up in Earth's atmosphere. We model the atmosphere with a fixed surface mixing ratio of CO$_2$ = 0.01 and O$_2$ = $2.1 \times 10^{-3}$ (1\% PAL = Present Atmospheric Level) and a stellar flux of 87\%. For the Earth-Sun case we fix the mixing ratios of the other biological gases to CH$_4$ = $7.07 \times 10^{-3}$ (4300x PAL), N$_2$O =  $8.37 \times 10^{-9}$, CH$_3$Cl = $5.0 \times 10^{-10}$ \citep[see][]{kaltenegger2007}. The corresponding biological surface fluxes required to sustain those mixing ratios are CH$_4$ = $1.01 \times 10^{17}$ g yr$^{-1}$, N$_2$O =  $3.14 \times 10^{13}$ g yr$^{-1}$, and CH$_3$Cl = $9.06 \times 10^{11}$ g yr$^{-1}$. All other stellar types except the Sun use these surface fluxes as the boundary conditions for those gases. For H$_2$ and CO we used fixed deposition velocities of $2.4 \times 10^{-4}$   and $1.2 \times 10^{-4}$ cm s$^{-1}$, respectively, corresponding to the maximum rate of transfer into an ocean \citep{domagal2014}. For the later stellar types, K7V, M1V, M3V and the three observed MUSCLES stars, the CH$_4$ boundary condition was changed to a fixed mixing ratio \citep[see also][]{segura2005, rugheimer2014}. For those stellar types, we used a mixing ratio of CH$_4$ corresponding to the last stable value, calculated for the K2V case, of $4.5 \times 10^{-3}$. For the M8V model, we needed to further reduce the CH$_4$ mixing ratio to $4.0 \times 10^{-3}$ for a stable atmosphere.

\begin{table}[ht!]
\begin{center}
\caption{Ozone Column Depths for the Grid of Host Stars (F0V to M8V) \label{tableOzone}}
\begin{tabular}{lccccccc}
\tableline\tableline

 & \multicolumn{4}{c}{O$_3$ Column Depth} \\
 \hline
 Star  & 3.9 Ga & 2.0 Ga & 0.8 Ga & Modern Earth  \\
  \hline   
\tableline
F0V        &  $1.76 \times 10^{16}$        &  $5.13 \times 10^{18}$&  $9.07 \times 10^{18}$  &  $1.15 \times 10^{19}$ \\
F7V      &  $3.93 \times 10^{15}$         &  $2.20 \times 10^{18}$ &  $6.17 \times 10^{18}$ &  $8.62 \times 10^{18}$ \\
Sun       &  $1.84 \times 10^{15}$       &  $6.01 \times 10^{17}$&  $4.23 \times 10^{18}$ &  $5.29 \times 10^{18}$ \\
G8V       &  $4.21 \times 10^{14}$      &  $1.73 \times 10^{17}$ &  $1.62 \times 10^{18}$ &  $5.56 \times 10^{18}$ \\
K2V       &  $7.32 \times 10^{14}$       &  $1.23 \times 10^{17}$&   $1.08 \times 10^{18}$ & $3.62 \times 10^{18}$ \\
K7V       &  $1.10 \times 10^{16}$       &  $5.31 \times 10^{17}$&    $7.38 \times 10^{17}$ & $3.46 \times 10^{18}$   \\
M1V       &  $1.01 \times 10^{16}$      &  $5.37 \times 10^{17}$ &  $7.37 \times 10^{17}$ & $ 4.03 \times 10^{18}$ \\
M3V       &  $9.88 \times 10^{15}$       &  $3.97 \times 10^{17}$&  $5.66 \times 10^{17}$ & $3.59 \times 10^{18}$   \\
M8V       &  $1.43 \times 10^{15}$       &  $2.63 \times 10^{17}$&   $3.15 \times 10^{17}$ &$2.47 \times 10^{18}$  \\
GJ581   &  $6.76 \times 10^{15}$       &  $3.55 \times 10^{17}$&  $4.75 \times 10^{17}$ & $1.18 \times 10^{18}$ \\
GJ832  &  $2.49 \times 10^{15}$       &  $3.46 \times 10^{17}$&   $6.25 \times 10^{17}$ &$1.59 \times 10^{18}$ \\
GJ1214&  $1.91 \times 10^{16}$      &  $6.29 \times 10^{17}$&  $9.00 \times 10^{17}$ &$ 2.31 \times 10^{18}$  \\

\tableline
\end{tabular}
\end{center}
\end{table}

The third epoch corresponds to Earth as oxygen continues to rise at 0.8 Ga and the start of the proliferation of multicellular life. We model the atmosphere with a fixed mixing ratio of CO$_2$ = 0.01 and O$_2$ = $2.1 \times 10^{-2}$ (10\% PAL) and a stellar flux of 94.8\%. For the Earth-Sun case we fix the mixing ratios of the other biological gases to CH$_4$ = $1.65 \times 10^{-3}$ (1000x PAL), N$_2$O =  $9.15 \times 10^{-8}$, CH$_3$Cl = $5.0 \times 10^{-10}$ (see \citet{kaltenegger2007}). The corresponding fluxes used for other stellar types are CH$_4$ = $2.75 \times 10^{16}$ g yr$^{-1}$, N$_2$O =  $2.08 \times 10^{13}$ g yr$^{-1}$, and CH$_3$Cl = $7.76 \times 10^{11}$ g yr$^{-1}$. For H$_2$ and CO we used fixed deposition velocities as above. For the later stellar types, M3V, M8V and the three observed MUSCLES stars, the CH$_4$ boundary condition was changed to a fixed mixing ratio. We used a mixing ratio of CH$_4$ corresponding to the last stable value, calculated for the M1V case, of $1.1 \times 10^{-2}$.

The fourth epoch corresponds to modern Earth. The model atmosphere has a fixed mixing ratio of CO$_2$ = 355ppm and O$_2$ = 0.21. The biogenic fluxes were held fixed in the models in accordance with the fluxes that reproduce the modern mixing ratios in the Earth-Sun case. The surface fluxes for long-lived gases H$_2$, CH$_4$, N$_2$O, CO and CH$_3$Cl were calculated such that the Earth around the Sun yields a $\tsurf$ = 288K for surface mixing ratios: cH$_2 = 5.5 \times 10^{-7}$, cCH$_4 = 1.6 \times 10^{-6}$, cCO$_2 = 3.5 \times 10^{-4}$, cN$_2$O $= 3.0 \times 10^{-7}$, cCO = $9.0 \times 10^{-8}$, and cCH$_3$Cl$ = 5.0 \times 10^{-10}$ \citep[see][]{rugheimer2013}. The corresponding surface fluxes are $-1.9 \times 10^{12}$ g H$_2$ yr$^{-1}$, $5.3 \times 10^{14}$ g CH$_4$ yr$^{-1}$, $7.9 \times 10^{12}$ g N$_2$O yr$^{-1}$, $1.8 \times 10^{15}$ g CO yr$^{-1}$, and $4.3 \times 10^{12}$ g CH$_3$Cl yr$^{-1}$. For the M8V, CH$_4$ and N$_2$O were given a fixed mixing ratio of $1.0\times 10^{-3}$ and $1.5\times 10^{-2}$ respectively \citep[following][]{rugheimer2014}.

\section{RESULTS: UV FLUXES FOR PLANETS ORBITING FGKM STARS}

\begin{figure*}[ht!]
\centering
\includegraphics[scale=0.5,angle=0]{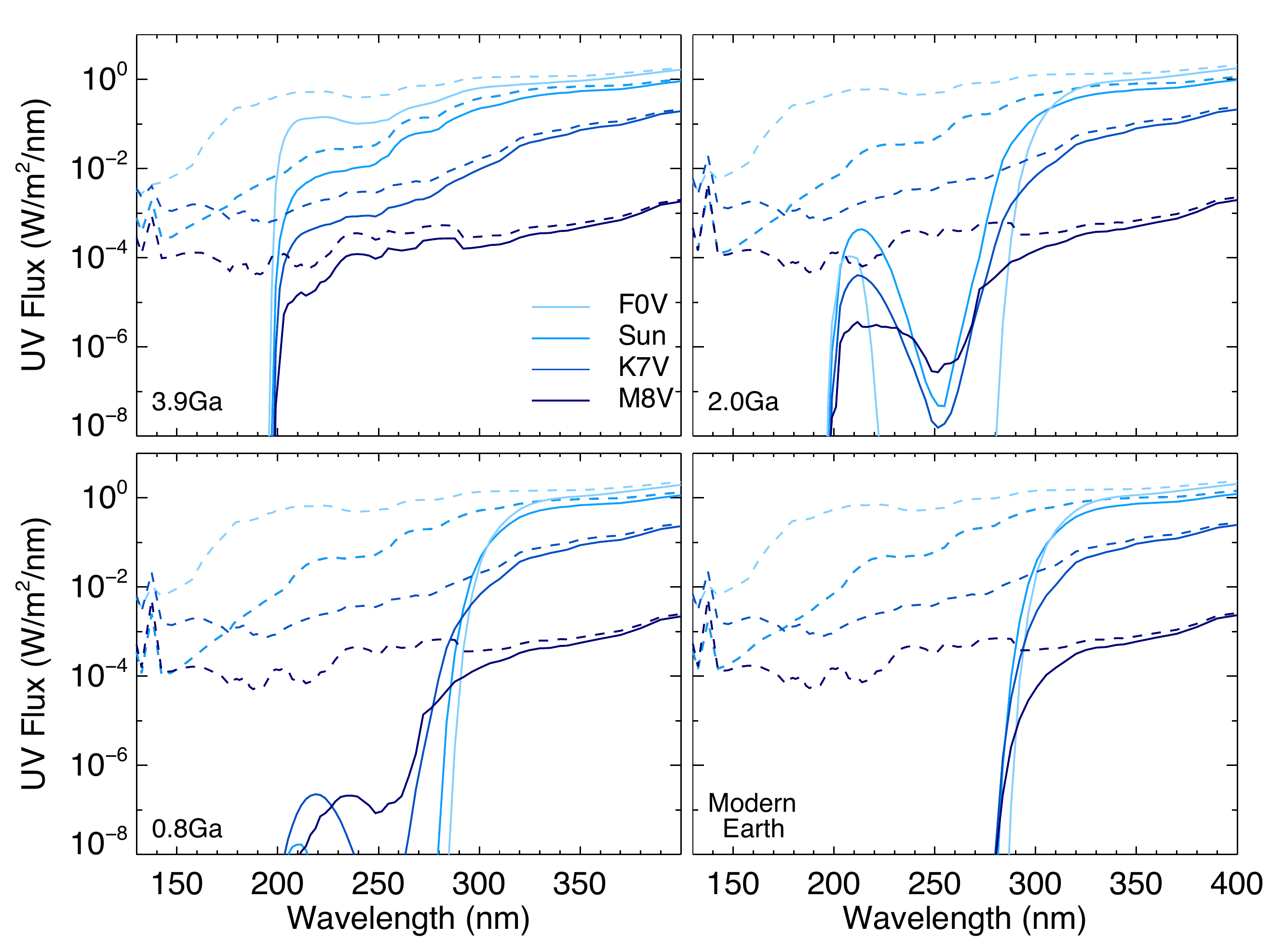}
\caption{The incident stellar UV flux (dashed lines) and the surface UV flux (solid lines) at four geological epochs for a sample of four stellar types spanning the stellar effective temperature range for FGKM stars. \label{Ep0}}
\end{figure*}

For each model atmosphere and star type we calculate the UV flux reaching the surface of an Earth-like planet and compare that to the UV flux incident on the top-of-atmosphere (TOA). In Fig. \ref{StarGround} we show the flux of the host star (black lines), the flux at the TOA (gray lines), and the flux reaching the ground for each epoch (colored lines) as seen from an observer at 10 pc. The flux from the host star relative to what is incident on the top of the atmosphere at 500nm ranges from 10$^3$ to 10$^{10}$ for the M8V to F0V, respectively.

Fig \ref{Ep0} compares the incident (dashed) and surface (solid) flux at the planet at each of the four geological epochs for an F0V ($\teff$=7000K), the Sun ($\teff$=5778K), a K7V ($\teff$=4250K), and an M8V star ($\teff$=2300K) as a representation of the twelve stars modeled. For a pre-biotic atmosphere, the UV surface environment on planets orbiting any of the grid stars follow a similar trend with TOA flux levels being similar to surface flux levels for $\lambda >$ 200nm (see top left panel Fig. \ref{Ep0}). At 200nm there is a sharp absorption in the atmosphere by CO$_2$ and H$_2$O and no UV flux at wavelengths below 200nm reaches the surface of the planet for any star type considered. With the rise of oxygen at 2.0 Ga to 1\% PAL we see absorption of UV photons around 250nm due to ozone (Fig. \ref{Ep0}, upper right). This is pronounced for F stars which have ozone column depths an order of magnitude higher at 2.0 Ga than G, K and M stars due to higher incident UV flux and photolysis of O$_2$ producing O$_3$ (see Table \ref{tableOzone} for O$_3$ column depths for all stars and epochs). As oxygen continues to rise at 0.8 Ga to 10\% PAL and then to current levels for the modern Earth, absorption increases for $\lambda <$ 280nm due to ozone, effectively blocking UVC radiation from reaching the surface (see also Table \ref{tableUVC} in appendix).

Table \ref{tableUV} lists the top-of-atmosphere (TOA) and surface UV fluxes for each stellar type and geological era for planets orbiting at the 1AU equivalent distance for their host star.  In the Appendix tables A1 to A3 show the TOA and surface UV fluxes subdivided into UVA (315-400nm), UVB (280-315nm), and UVC (121.6-280nm). Note that all fluxes have been diurnally averaged and calculated for a zenith angle of 60 degrees, representing a global average. We discuss the F0V-M8V results below and, comparable information for the three MUSCLES stars can be found in Table \ref{tableUV} and Tables A1-A3.

At 3.9 Ga, 54\% - 77\% of the total incoming stellar UV flux (121.6nm to 400nm) reaches the surface for a pre-biotic atmosphere orbiting different host stars. The lowest percentage of UV radiation reaches the ground for a pre-biotic Earth orbiting our M8V model star, and the highest percentage of UV radiation reaches the ground a K7V star (this trend holds for all epochs). Note that this does not correspond to the min and max of the absolute transmitted flux. The minimum UV flux that reaches the ground is 0.035 W m$^{-2}$ for an M8V host star and the maximum is 61 W m$^{-2}$ for an F0V host star in this epoch. However, the fraction of UVA, UVB, and UVC fluxes reaching the surface is different. In the UVA region, between 80\% (G2V) - 82\% (M8V) of the incident flux and between 0.026 W m$^{-2}$ (M8V) and 45 W m$^{-2}$ (F0V) reaches the surface. In the UVB region, between 57\% (F0V) and 59 \% (K2V) of the incident flux and between 0.0030 W m$^{-2}$ (M8V) and 10 W m$^{-2}$ (F0V) reaches the surface. In the UVC region between 15\% (M1V) and 43\% (G8V) of the incident flux and between $4.2 \times 10^{-3}$ W m$^{-2}$ (M8V) and 5.9 W m$^{-2}$ (F0V) reaches the surface. The percentage and absolute flux reaching the ground depends on the interplay between the intensity of incoming UV radiation and the amount of ozone shielding.

At 2.0 Ga, 25\% (M8V) - 69\% (K7V) of the total incoming stellar UV flux reaches the surface. In absolute flux, the minimum UV flux that reaches the surface is 0.032 W m$^{-2}$ (M8V) and the maximum is 48 W m$^{-2}$ (F0V). In the UVA region, between 73\% (Sun) - 78\% (K7V) of the incident flux and between 0.029 W m$^{-2}$ (M8V) and 47 W m$^{-2}$ (F0V) reaches the surface for a planet. In the UVB region, between 7.8\% (F0V) and 48\% (K2V) of the incident flux and between  $2.2 \times 10^{-3}$ W m$^{-2}$ (M8V) and 2.5 W m$^{-2}$ (G8V) reaches the surface. In the UVC region between $2.1 \times 10^{-3}$ \% (F0V) and 2.5\% (G8V) of the incident flux and between $1.7 \times 10^{-4}$ W m$^{-2}$ (M8V) and $5.8 \times 10^{-4}$ W m$^{-2}$  (F0V) reaches the surface for a planet. 

At 0.8 Ga, 26\% (M8V) - 69\% (K7V) of the total incoming stellar UV flux reaches the surface. In absolute flux, the minimum UV flux that reaches the surface is 0.035 W m$^{-2}$ (M8V) and the maximum is 51 W m$^{-2}$ (F0V). In the UVA region, between 72\% (F0V) - 79\% (M8V) of the incident flux and between 0.032 W m$^{-2}$ (M8V) and 50 W m$^{-2}$ (F0V) reaches the surface for a planet. In the UVB region, between 4.0\% (F0V) and 36\% (M8V) of the incident flux and between  $2.3 \times 10^{-3}$ W m$^{-2}$ (M8V) and 1.0 W m$^{-2}$ (F7V) reaches the surface. In the UVC region between $1.3 \times 10^{-9}$\% (F0V) and 0.097\% (M8V) of the incident flux and between $4.1 \times 10^{-8}$ W m$^{-2}$ (F0V) and $1.4 \times 10^{-4}$ W m$^{-2}$  (M3V) reaches the surface for a planet. 

For the modern Earth-like atmosphere, 25\% (M8V) - 69\% (K7V) of the total incoming stellar UV flux reaches the surface. In absolute flux, the minimum UV flux that reaches the surface is 0.034 W m$^{-2}$ (M8V) and the maximum is 53 W m$^{-2}$ (F0V). In the UVA region, between 72\% (F0V) - 79\% (M8V) of the incident flux and between 0.034 W m$^{-2}$ (M8V) and 53 W m$^{-2}$ (F0V) reaches the surface for a planet. In the UVB region, between 2.7\% (F0V) and 17\% (K7V) of the incident flux and between $9.9\times 10^{-4}$ W m$^{-2}$ (M8V) and 0.80 W m$^{-2}$ (Sun) reaches the surface. In the UVC region between $7.7 \times 10^{-29}$\% (F0V) and  $4.6 \times 10^{-9}$\% (M8V) of the incident flux and between $2.5\times 10^{-29}$ W m$^{-2}$ (F0V) and $4.3 \times 10^{-12}$ W m$^{-2}$  (M8V) reaches the surface for a planet.


\begin{table*}[h!]
\begin{center}
\caption{Top-of-Atmosphere (TOA) and Surface UV fluxes for F G K \& M Stars \label{tableUV}}
\begin{tabular}{llllllllllllllllll}
\tableline\tableline

& \multicolumn{8}{c}{UV fluxes 121.6 to 400 nm (W m$^{-2}$)} \\
 \hline
 Star  & \multicolumn{2}{c}{3.9 Ga} & \multicolumn{2}{c}{2.0 Ga} &   \multicolumn{2}{c}{0.8 Ga} & \multicolumn{2}{c}{Modern Earth}   \\
Type   & TOA & Surface & TOA & Surface & TOA & Surface & TOA & Surface  \\
  \hline   
\tableline
F0V   &    98.6  &  60.8  &  113.  &  48.1  &  123. & 51.1    & 130. & 53.3  \\
F7V    &   62.6 &  44.2   &  71.7  &  39.1  &  78.1  & 40.8  &  82.4 & 42.5  \\
Sun    &   37.9 &  28.1   &  43.6  &  27.8  &  51.4  &  30.4  & 55.0  &  32.3  \\
G8V   &   31.3 &  23.4   &  35.8   & 23.8   &  39.0  & 24.4   &  41.2 & 25.0  \\
K2V    &  16.5 &  12.6   &  19.0   & 13.1  &  20.7   &  13.8  &  21.8  & 14.2   \\
K7V    &  4.79  &  3.71  &  5.68   & 3.94   &  6.19   & 4.29  & 6.53 & 4.50  \\
M1V   &   2.00  &  1.22  &  3.39   & 1.19  &  3.69   & 1.30 & 3.89 & 1.36   \\
M3V   &   0.934 &  0.510  &  1.75   & 0.465  &  1.91  & 0.503 &  2.01 & 0.533 \\
M8V   &   0.0646 & 0.0347 & 0.125 &  0.0316 & 0.136 & 0.0349 & 0.144 & 0.0353 \\
GJ 581   &  1.57 &  1.25  &  2.04 &  1.35 &  2.04    &  1.47  & 2.15 & 1.57    \\
GJ 832 &  2.06 & 1.61    &  2.64  &  1.74 &  2.87   &  1.90  &  3.03 & 2.02   \\
GJ 1214 & 0.943 &  0.722  &  1.19  &  0.778 &  1.19  &  0.847  & 1.25 & 0.902  \\
\tableline
\end{tabular}
\end{center}
\end{table*}

\begin{figure*}[h!]
\centering
\includegraphics[scale=0.4,angle=0]{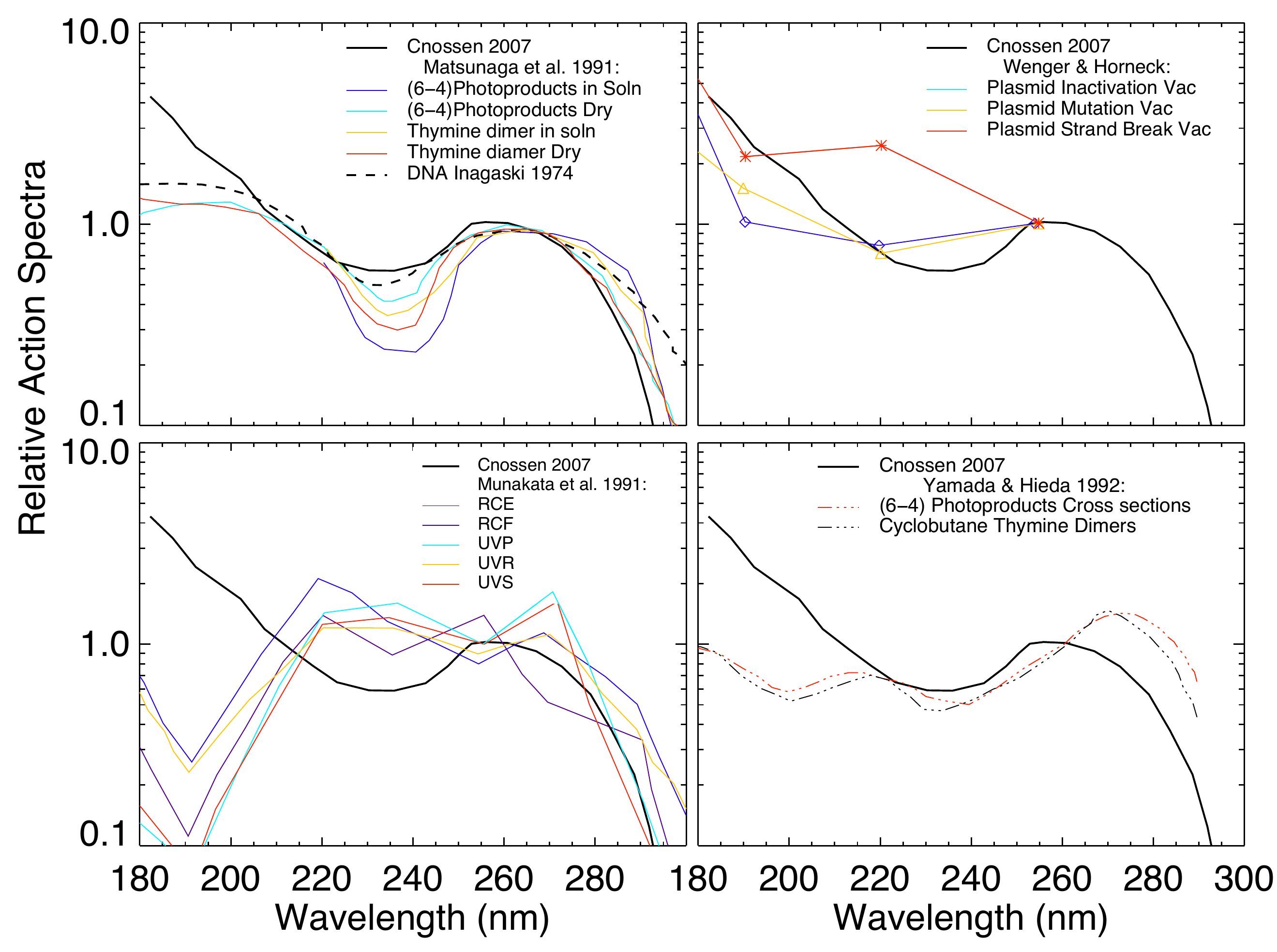}
\caption{Action Spectra for photoproduct formation (top - left, bottom - right), DNA plasmid inactivation, mutation, and strand-breaks (top - right), and the inactivation of bacterial spores (bottom - left).\label{comparingaction}}
\end{figure*}

\begin{figure*}[ht!]
\centering
\includegraphics[scale=0.5,angle=0]{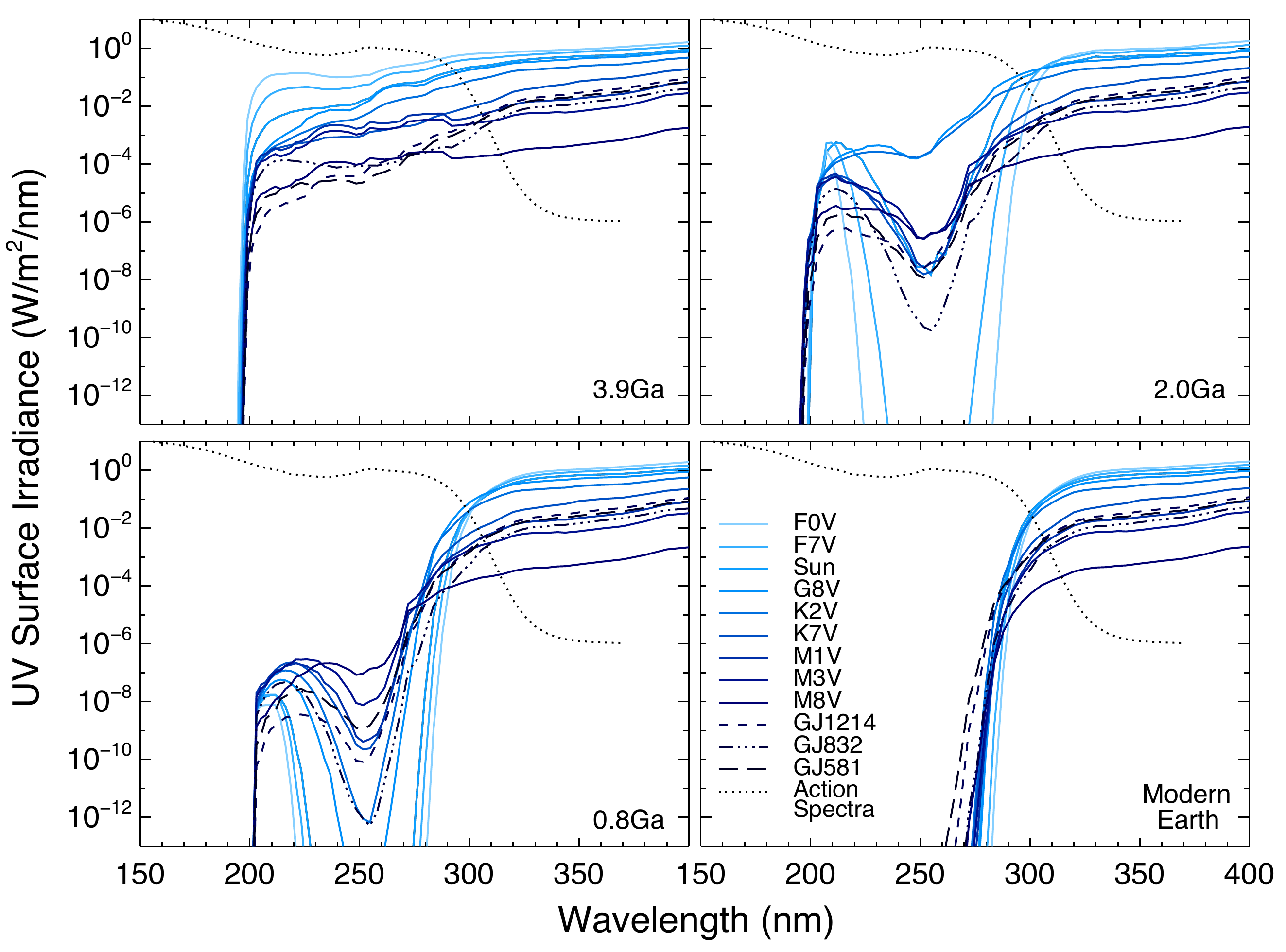}
\caption{Surface UV flux levels (W m$^{-2}$ nm$^{-1}$) for an Earth-sized planet at four geological epochs orbiting F0V to M8V host stars. Lines have been smoothed for clarity. DNA action spectrum is shown as a dotted line. \label{EpochsUV}}
\end{figure*}

To estimate the biologically relevant UV fluxes, it is useful to consider the damage to DNA and other biomolecules. An action spectrum is a parameter which gives the relative biological response effectiveness at different wavelengths. Multiplying the surface flux by the action spectrum gives the biologically effective irradiance. Integrated for a wavelength interval gives the biochemical effectiveness in that region, $\Eeff$:

\begin{equation}
\Eeff = \int_{\lambda_1}^{\lambda_2} F_\mathrm{surf}(\lambda) S_\lambda(\lambda) d\lambda
\end{equation}

\noindent where $F_\mathrm{surf}(\lambda)$ is the surface flux (W m$^{-2}$ nm$^{-1}$) and $S_\lambda(\lambda)$ is the action spectrum of the biomolecule of interest in relative units. Action spectra are typically normalized at 260 nm or 300 nm and are given in relative units. The damage of such molecules as the thymine dimers and (6-4)photoproducts correspond to DNA damage and have similar action spectra \citep[see Fig. \ref{comparingaction} and][]{matsunaga1991}. We use a DNA action spectrum for 182 nm $< \lambda < 370\ $ nm based on \citet{cnossen2007} for $\lambda < 260$ nm and \citet{setlow1974}\footnote{From database: http://www.esrl.noaa.gov/gmd/grad/antuv/docs/version2/descVersion2Database3.html} for $\lambda > 265$ nm.  For comparison to other shortwave action spectra ($ \lambda < 300\ $) we plot the response curves for DNA photoproducts formation \citep{matsunaga1991, yamada1992}, DNA plasmid inactivation, mutation and strand-breaks \citep{wehner1995}, and the inactivation of bacterial spores \citep{munakata1991} (see Fig. \ref{comparingaction}). The action spectra agree to within an order of magnitude and is shown to 360nm in Fig. \ref{EpochsUV}.

We normalized our DNA action spectrum at 260\ nm and reference to modern Earth-Sun when quoting the biologically effective irradiance as given in Equation \ref{Eeff} to estimate the damage through geological time.

\begin{equation}\label{Eeff}
\Eeff =  \frac{ \int_{182\ nm}^{370\ nm} F_\mathrm{surf}(\lambda) S_{\lambda\ DNA}(\lambda) d\lambda}{   \int_{182\ nm}^{370\ nm} F_\mathrm{surf\ modern\ \earth}(\lambda) S_{\lambda\ DNA}(\lambda) d\lambda}
\end{equation}

Table \ref{tableEeff} gives the biologically effective irradiance and the absolute UV flux ratio compared to the modern Earth-Sun model for each geological epoch and stellar type.

\begin{table*}[ht!]
\begin{center}
\caption{Biologically effective irradiance $\Eeff$, relative to Modern Earth levels\label{tableEeff}}
\begin{tabular}{llllllllllllllllll}

\tableline\tableline
 
& \multicolumn{8}{c}{Fluxes integrated from 182nm to 370nm} \\
 \hline
 Star  & \multicolumn{2}{c}{3.9 Ga} & \multicolumn{2}{c}{2.0 Ga} &   \multicolumn{2}{c}{0.8 Ga} & \multicolumn{2}{c}{Modern Earth}   \\
Type   & $\Eeff$ & UV Ratio & $\Eeff$ & UV Ratio  & $\Eeff$ & UV Ratio  & $\Eeff$ & UV Ratio   \\
  \hline   
\tableline
F0V   &    3520 &  2.19  &  2.75  &  1.42  &  0.655 & 1.47    & 0.367  & 1.51  \\
F7V  &   1680 & 1.57  &  10.8  & 1.23   &   1.13 & 1.23   & 0.524  & 1.27  \\
Sun    &   611 &  0.969  &  41.3  &  0.906  &  1.53  &  0.942  & 1.00  &  1.00  \\
G8V   &  482  & 0.806  &   107 &  0.797  &  7.71  & 0.781   & 0.678  & 0.773 \\
K2V    &   181 & 0.416  &     56.5 &  0.425  & 5.93   &  0.433 & 0.671  & 0.436 \\
K7V    &  26.2 &  0.102    &  1.95   & 0.105  & 1.28 & 0.114 & 0.105 & 0.117 \\
M1V    &  44.5  & 0.0345  & 1.13   &  0.0301  &  0.662  & 0.0327   &  0.0317 & 0.0330  \\
M3V    & 27.8   & 0.0150  &  1.26   &  0.0118  &  0.644  &   0.0126 &  0.0245 & 0.0126  \\
M8V   &   2.33 & 0.00104 & 0.182 &  0.000807  & 0.132 & 0.000881 & 0.00393 & 0.000839 \\
GJ 581  &   2.02 &  0.0312 &  0.432  &  0.0335  &  0.345  &  0.0364  & 0.120  & 0.0385 \\
GJ 832   &   3.37 & 0.0403  &  0.616  &  0.0431  &  0.353  & 0.0468   & 0.105  & 0.0493  \\
GJ 1214   &   2.67 &  0.0176 &  0.153  & 0.0186   &  0.0693  &  0.0202  &  0.0180 & 0.0213 \\
\tableline
\end{tabular}
\end{center}
\end{table*}

\begin{figure}[h!]
\centering
\includegraphics[scale=0.45,angle=0]{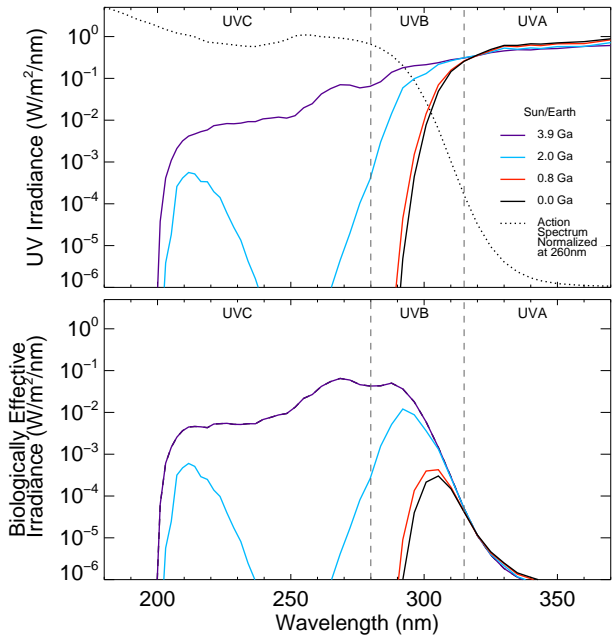}
\caption{UV fluxes at the surface for the Earth-Sun case through time with the dashed line as the DNA Damage Action Spectrum (top). The relative biological effectiveness of UV damage is the convolution of the flux and action spectrum over wavelength (bottom).\label{ActionDNA}}
\end{figure}

Because biological damage is heavily focused in the UVC and UVB regions (shown in Fig. \ref{ActionDNA}, the $\Eeff$ values are much higher for the early Earth case than is intuitive from the absolute fluxes, especially around F-type host stars. For an F0V prebiotic atmosphere, we calculate $\Eeff$ = 3520, three orders of magnitude larger than for modern Earth and 6 times larger than the early Earth-Sun model. This $\Eeff$ value is 1600 times larger than the ratio of UV fluxes at the surface for a prebiotic world orbiting an F0V host star, which highlights the importance of including the action spectrum when considering biological effects. For the anoxic atmosphere at 3.9 Ga, all stars have $\Eeff$ larger than one, the modern Earth-Sun value. We find the smallest $\Eeff$ values are about 2 times the modern Earth-Sun value in our model for the surface of a prebiotic planet orbiting GJ 581. It has 31 times less total UV ground radiation and 300 times less biologically effective radiation as the prebiotic Earth-Sun model.
\newpage
At the rise of oxygen at 2.0 Ga (1\% PAL O$_2$), we see a substantial reduction in $\Eeff$ values. The surface of planets orbiting F, G, and K grid stars still receive UV radiation corresponding to  $\Eeff$ values higher than modern Earth-Sun, but planets orbiting M stars receive biologically effective irradiances near or less than modern Earth. At 0.8 Ga (10\% PAL O$_2$), we see further reduction in $\Eeff$ values. For modern Earth's atmosphere composition, the surface of planets orbiting all other host stars receive UV fluxes that correspond to $\Eeff$ values less than one. For the F stars, this is due to increased shielding from higher abundances of ozone. For the cooler K and M stars, this is due to a lower abundance of stellar UV photons incident on the atmosphere.

In Fig. \ref{ActionDNA} we show the action spectrum (normalized at 260 nm, dashed line) with the Earth-Sun UV surface fluxes through the four geological epochs. Note the strong overlap for the pre-biotic atmosphere. Already at 2.0 Ga most of the UV flux has been attenuated by O$_3$ in the atmosphere, causing a sharp decrease in the $\Eeff$ values for the later epochs. The bottom panel of Fig. \ref{ActionDNA} shows the surface UV flux weighted by the DNA action spectrum, highlighting which wavelengths and epochs have the most damaging irradiation for molecules with similar responses as the DNA action curve.

Despite harsh UVB and UVC radiation conditions present on a planet without an ozone shield, certainly prebiotic chemistry and the origin of life flourished on the early Earth. Being under a layer of water or rock would mitigate the problems induced by high UV fluxes. Microbial mats scatter light such that the lower levels have only 1\% of incident light at 0.5mm depth, and subsurface community hosting layers of sandstone reduce light levels to 0.005\% of incidence \citep{cockell1999, garcia1994, nienow1993, nienow1988}. As well, prebiotic organic polymers and dissolved inorganic ions may provide sufficient protection from UV degradation in as little as 2mm of ocean water \citep{cleaves1998}.

Fig. \ref{EpochsUV} shows the smoothed lines for the surface UV flux for an Earth-sized planet at the four geological epochs with the action spectrum curve overlaid for all grid stars modeled. 

\section{DISCUSSION \& CONCLUSIONS}

In this study, we investigated the exoplanet surface UV fluxes for Earth-like planets orbiting a grid of F, G, K, and M stars at four possible biogeological epochs in the evolution of life. We use the input flux from 12 different stellar types, nine model stars in the FGKM range with $\teff$ = 4250K to $\teff$ = 7000K, and three M stars (GJ 581, GJ 832, and GJ 1214) with observed UV fluxes for our calculations. The atmospheres selected correspond to a pre-biotic Earth model at 3.9 Ga, a model atmosphere for the early rise of oxygen at 2.0 Ga and 0.8 Ga, and modern Earth biology. In terms of total UV flux at the surface, an F star receives 1-2 times more flux than the Sun, where as an M star receives 20-900 times less flux than the Sun depending on the sub-spectral class. 

We also consider the amount of UVA, UVB and UVC fluxes, with the latter two showing the greatest changes among stellar types and geological epochs. For a pre-biotic atmosphere, a significant portion of incident UVC flux reaches the surface. Whereas in the modern atmosphere for planets orbiting any grid star, all UVC photons are effectively shielded by ozone. For all epochs, the smallest percentage of total UV flux compared to incident flux reaching the ground is for an M8V star and the highest fraction is for a K7V star due to the balance of available near- and far-UV photons and the amounts of ozone produced. However in terms of absolute flux, the smallest amount reaches the surface is for an M8V host star and the largest amount for an F0V star.

To analyze the potential biological consequences of the UV surface environment for the origin and evolution of life, we convolve the UV surface fluxes with the DNA action spectrum as a proxy for biological damage. While it is uncertain whether DNA will arise as the primary information carrier on another planet, current biosignature searches are focused on finding carbon-based chemistry. Since the action spectra for other carbon based biomolecules follow similar trends in their actions spectra, we use DNA as a proxy for biochemical effectiveness of UV radiation damage. The surface of pre-biotic worlds have significantly higher $\Eeff$ values than around the early Sun. \emph{Deinococcus radiodurans} is one of the most radiation tolerant organism known, withstanding 1000x the lethal human radiation dose. UV tolerance experiments show \emph{D. radiodurans} can survive 400 J m$^{-2}$ of 254nm radiation without significant loss of viability with appreciable loss of viability occurring around 500-600 J m$^{-2}$ \citep{gascon1995}.  

We find that a pre-biotic Earth (3.9 Gyr ago) orbiting an F0V star receives 6 times the biologically effective radiation as around the early Sun and 3500 times the modern Earth-Sun levels. A pre-biotic Earth orbiting an M8V star around receives 300 times less biologically effective radiation and 2 times modern Earth-Sun levels. 

For the second geological epoch with 1\% PAL O$_2$, a planet orbiting an F0V and M8V star receives 15 and 230 times less biologically effective radiation as the Earth-Sun model at that geological epoch, respectively. For the third geological epoch with 10\% PAL O$_2$, a planet orbiting an F0V and M8V star receives 2 and 12 times less biologically effective radiation as the Earth-Sun model at that geological epoch, respectively. For the fourth geological epoch, corresponding to modern Earth concentrations of O$_2$, a planet orbiting an F0V and M8V star receives 3 and 250 times less biologically effective radiation as the modern Earth-Sun, respectively. Note that for all epochs after oxygen begins to rise in the atmosphere, both the hottest stars and coolest stars have less biologically effective radiation. For the hottest stars, this is due to increased ozone shielding from higher UV environments, and for the coolest stars this is due to less absolute UV flux.

A layer of water or soil might be needed to shield life for hot stars. Even for a planet orbiting the least UV active star in our grid, the biologically effective irradiance on the surface of a pre-biotic world is several times the value for modern Earth. With higher ozone concentration in the atmosphere, that value decreases for planets orbiting all star types. 

We model clear sky atmospheres for all planets at all epochs which on average will represent an upper limit to the UV radiation environment at the surface since clouds most frequently block up to 80\% of incoming UV radiation \cite{grant1997, parisi2004}. However clouds also can focus more UV radiation to the ground depending on cloud cover and type, increasing the amount of UV radiation reaching the surface. Earth measurements of UVA and UVB radiation on cloudy versus cloud-free days show clouds can enhance UV radiation by a factor of 1.08 and 1.4 for UVA and UVB radiation, respectively \citep{parisi2004}.

Note that higher CO$_2$ and CH$_4$ concentrations in the atmosphere as well as hazes could further protect the surface from UV radiation \citep{cnossen2007, cockell1999}. In addition, UV photons may also positively contribute to pre-biotic reactions leading to the origin of life \citep{senanayake2006, barks2010,sutherland2012, sutherland2015}. 

Our models provide boundary conditions for the UV environment on the surface of Earth-like planets orbiting a grid of FGKM stars from pre-biotic to modern Earth atmosphere. Our results inform  photo biological assessments, pre-biotic chemistry, and early life evolution experiments.

\section*{ACKNOWLEDGMENTS}

We would like to thank Kevin France for discussions concerning the MUSCLES database and Sukrit Ranjan for discussions concerning DNA action spectra. This work has made use of the MUSCLES M dwarf UV radiation field database. We would also like to acknowledge support from DFG funding ENP KA 3142/1-1 and the Simons Foundation (290357, Kaltenegger and 290360, Sasselov).

\section*{ABBREVIATIONS} 
Ga - Gyr ago or Billion years ago, TOA - Top of Atmosphere, PAL - Present Atmospheric Level

\newpage
\section*{Appendix}


\setcounter{table}{0}
\renewcommand{\thetable}{A\arabic{table}}

\begin{table*}[h!]
\begin{center}
\caption{Top-of-Atmosphere (TOA) and Surface UVA fluxes for F G K \& M Stars \label{tableUVA}}
\begin{tabular}{lllllllllllllllll}
\tableline\tableline
& \multicolumn{8}{c}{UVA fluxes 315 to 400 nm (W m$^{-2}$)} \\
 \hline
 Star  & \multicolumn{2}{c}{3.9 Ga} & \multicolumn{2}{c}{2.0 Ga} &   \multicolumn{2}{c}{0.8 Ga} & \multicolumn{2}{c}{Modern Earth}   \\
Type   & TOA & Surface & TOA & Surface & TOA & Surface & TOA & Surface  \\

  \hline   
\tableline
F0V        &  56.0  &  44.7  &  64.1  &  46.5  &  69.8  &  50.2   &  73.6  &  52.7  \\
F7V        &  43.8  &  34.9  &  50.2  &  37.0  &  54.7  &  39.8   &  57.7  &  41.8  \\
Sun        &  29.9  &  23.8  &  34.4  &  25.6  &  40.2  &  29.5   &  43.1  &  31.5  \\
G8V       &  25.0  &  19.9  &  28.6  &  21.3  &  31.1  &  23.0   &  32.8  &  24.4  \\
K2V        & 13.9  &  11.1  &  15.9  &  12.0  &  17.4  &  13.0   &  18.3  &  13.8  \\
K7V        & 4.31  &  3.51  &  4.94  &  3.83  &  5.38  &  4.18   &  5.68  &  4.45  \\
M1V       &  1.30  &  1.06  &  1.49  &  1.14  &  1.62  &  1.25   &  1.71  &  1.34  \\
M3V       & 0.505 &  0.412  &  0.578  &  0.435  &  0.630  &  0.475   & 0.664   &  0.521   \\ 
M8V       &0.0331 &  0.0260  &  0.0379  &  0.0292  &  0.0412  & 0.0324    & 0.0435     &  0.0343  \\
GJ 581   & 1.48  &  1.21  &  1.70  &  1.32  &  1.85  &  1.44   &  1.95  &  1.54  \\
GJ 832   & 1.91  &  1.56  &  2.19  &  1.70  &  2.38  &  1.86   &  2.51  & 1.98  \\
GJ 1214 & 0.858  &  0.700  &  0.983  &  0.764  &  1.07  &  0.832   &  1.13  &   0.890  \\

\tableline
\end{tabular}
\end{center}
\end{table*}

\begin{table*}[h!]
\begin{center}
\caption{Top-of-Atmosphere (TOA) and Surface UVB fluxes for F G K \& M Stars \label{tableUVB}}
\begin{tabular}{lllllllllllllllll}
\tableline\tableline
& \multicolumn{8}{c}{UVB fluxes 280 to 315 nm (W m$^{-2}$)} \\
 \hline
 Star  & \multicolumn{2}{c}{3.9 Ga} & \multicolumn{2}{c}{2.0 Ga} &   \multicolumn{2}{c}{0.8 Ga} & \multicolumn{2}{c}{Modern Earth}   \\
Type   & TOA & Surface & TOA & Surface & TOA & Surface & TOA & Surface  \\

  \hline   
\tableline
F0V        &  18.0  &   10.3  &  20.6  &   1.60  &   22.5  &    0.905    &    23.7  &   0.644   \\
F7V        &  11.4  &   6.67  &  13.1  &  2.13  &   14.2  &   0.998  &   15.0  &    0.692   \\
Sun        &  5.77  &   3.41  &  6.76  &  2.19  &   8.04  &   0.902  &   8.63  &    0.798   \\
G8V       &  4.71  &   2.79  &  5.39  &  2.45  &   5.87  &   1.31  &   6.19  &    0.583   \\
K2V        &  2.00  &   1.19  &  2.29  &  1.10  &   2.49  &   0.714  &   2.63  &    0.404   \\
K7V        &  0.261  &  0.153   & 0.145   &   0.110  &  0.325   &  0.108  & 0.343    &  0.0566     \\ 
M1V       &  0.126  &  0.0730   &  0.0844   &  0.0466  &   0.158  &   0.0461  &   0.266  &    0.0219   \\
M3V       &  0.0737  &  0.0422   &  0.00590  &  0.0281  &   0.0920  &   0.0275  &   0.0970  &    0.0120   \\
M8V       & 0.00516 & 0.00298   & 0.0863   &  0.00219  &   0.00643  &   0.00230  &   0.00678  &  0.000986   \\  
GJ 581   & 0.0578  &  0.0350   &  0.0721  &  0.0310  &    0.0721  &   0.0321  &   0.0760  &    0.0280   \\
GJ 832   & 0.0754  &  0.0457   &  0.0863  &  0.0399  &   0.0941  &   0.0389  &  0.0992  &    0.0322   \\
GJ 1214 & 0.0279  & 0.0169    & 0.0319   & 0.0141    &  0.0348  &   0.0143  &    0.0367  &    0.0116   \\

\tableline
\end{tabular}
\end{center}
\end{table*}

\begin{table*}[h!]
\begin{center}
\caption{Top-of-Atmosphere (TOA) and Surface UVC fluxes for F G K \& M Stars \label{tableUVC}}
\begin{tabular}{lllllllllllllllll}
\tableline\tableline
& \multicolumn{8}{c}{UVC fluxes 121.6 to 280 nm (W m$^{-2}$)} \\
 \hline
 Star  & \multicolumn{2}{c}{3.9 Ga} & \multicolumn{2}{c}{2.0 Ga} &   \multicolumn{2}{c}{0.8 Ga} & \multicolumn{2}{c}{Modern Earth}   \\
Type   & TOA & Surface & TOA & Surface & TOA & Surface & TOA & Surface  \\
 
  \hline   
\tableline
F0V   &    24.7  &  5.89 $\times 10^{0}$  &  28.4 & $5.83 \times 10^{-4}$  & 30.9  & $4.08 \times 10^{-8}$  & 32.6 & $2.50 \times 10^{-29}$ \\
F7V    &   7.46  &  2.64 $\times 10^{0}$  & 8.57 & $2.00 \times 10^{-3}$   & 9.34  &$9.42 \times 10^{-8}$  & 9.86 & $1.63 \times 10^{-23}$ \\
Sun    &   2.20  &  8.71 $\times 10^{-1}$ & 2.63 & $4.03 \times 10^{-3}$   &  3.20  & $9.31 \times 10^{-8}$  & 3.38 & $2.30 \times 10^{-16}$ \\
G8V   &   1.59 &  6.86 $\times 10^{-1}$ & 1.84 & $4.58 \times 10^{-2}$   &  2.01  & $5.58 \times 10^{-7}$  & 2.12 & $3.96 \times 10^{-17}$ \\
K2V    &  0.589 &  2.46 $\times 10^{-1}$ &  0.709 & $3.48 \times 10^{-2}$   & 0.773  &$7.79 \times 10^{-6}$  & 0.814 & $4.17 \times 10^{-13}$ \\
K7V    &  0.209 & 3.99 $\times 10^{-2}$ & 0.441 &  $3.88 \times 10^{-4}$   & 0.480  &$1.39 \times 10^{-5}$  & 0.506 & $1.14 \times 10^{-13}$ \\
M1V   &   0.549 & 7.97 $\times 10^{-2}$ & 1.73 & $6.14 \times 10^{-4}$   & 1.88  &$5.79 \times 10^{-5}$  &  1.99  & $4.57 \times 10^{-15}$  \\
M3V   &   0.342 & 4.98 $\times 10^{-2}$ &  1.07 & $8.55 \times 10^{-4}$   & 1.17  &$1.39 \times 10^{-4}$  &  1.23 & $3.38 \times 10^{-14}$  \\
M8V   &   0.0254 & 4.22 $\times 10^{-3}$ & 0.0802 &   $1.72 \times 10^{-4}$   & 0.0874  & $8.49 \times 10^{-5}$  & 0.0921 & $4.27 \times 10^{-12}$  \\ 
GJ 581   & 0.0253 & 2.03 $\times 10^{-3}$ & 0.110 &  $6.25 \times 10^{-5}$   & 0.119  &$1.75 \times 10^{-5}$  &  0.126 & $6.87 \times 10^{-8}$ \\
GJ 832 & 0.0706 &  3.77 $\times 10^{-3}$ & 0.360 &  $9.99 \times 10^{-5}$   & 0.392  &$1.01 \times 10^{-5}$   & 0.414  & $6.08 \times 10^{-9}$ \\
GJ 1214 & 0.0559 & 4.63 $\times 10^{-3}$ & 0.0742 &  $1.00 \times 10^{-4}$   & 0.0808  &$1.00 \times 10^{-6}$  &  0.0853 & $1.15 \times 10^{-11}$  \\

\tableline
\end{tabular}
\end{center}
\end{table*}


\end{document}